\documentclass[english]{IEEEtran}
\usepackage[T1]{fontenc}

  \usepackage{nopageno}
\usepackage[latin9]{inputenc}

\usepackage{url}

\PassOptionsToPackage{draft}{hyperref}
\usepackage[bookmarks=false]{hyperref}
\PassOptionsToPackage{bookmarks=false}{hyperref}

\usepackage{color}
\usepackage{xcolor}
\usepackage{array}
\usepackage{booktabs}
\usepackage{textcomp}
\usepackage{multirow}
\usepackage{amsmath}
\usepackage{amsthm}
\usepackage{amssymb}
\usepackage[export]{adjustbox}
\usepackage{verbatim}
\usepackage{graphicx}
\usepackage{booktabs}

\newcounter{defcounter}
\usepackage{caption}
\usepackage[shortlabels]{enumitem}
\usepackage{cite}
\usepackage[ruled]{algorithm2e}
\mathchardef\period=\mathcode`.
\DeclareMathSymbol{.}{\mathord}{letters}{"3B}
\usepackage{tikz}
\tikzstyle{io} = [fill=black,inner sep=2pt,circle]

\graphicspath{ {images/} }

\usetikzlibrary{arrows,positioning}

\everymath{\displaystyle}
\usetikzlibrary{backgrounds}

\usepackage{pifont}

\makeatletter
\def\endthebibliography{%
	\def\@noitemerr{\@latex@warning{Empty `thebibliography' environment}}%
	\endlist
}

\makeatletter
\newcommand*\bigcdot{\mathpalette\bigcdot@{.5}}
\newcommand*\bigcdot@[2]{\mathbin{\vcenter{\hbox{\scalebox{#2}{$\m@th#1\bullet$}}}}}
\makeatother

\makeatletter

\theoremstyle{plain}

\usepackage[caption=false,font=normalsize,labelfont=sf,textfont=sf]{subfig}
\usepackage{cite} 
\usetikzlibrary{shapes,arrows}
\tikzstyle{line}=[draw] 
\usepackage{algorithmic}
\usepackage{amsmath,varwidth,array,ragged2e}
\usepackage{graphicx}
\usepackage{enumitem}
\usepackage{amsfonts}
\usepackage{mathrsfs}

\makeatother
\usepackage{babel}
\providecommand{\theoremname}{Theorem}

\begin{document}

\title{Blockchain-based Charging Coordination Mechanism for Smart Grid Energy Storage Units}
	
\author{Mohamed~Baza\IEEEauthorrefmark{1},~Mahmoud~Nabil\IEEEauthorrefmark{1},~Muhammad Ismail\IEEEauthorrefmark{2},~Mohamed~Mahmoud\IEEEauthorrefmark{1},~Erchin Serpedin\IEEEauthorrefmark{2},~and Mohammad Ashiqur Rahman\IEEEauthorrefmark{3}
		\\\IEEEauthorblockA{\IEEEauthorrefmark{1}Department of Electrical and Computer Engineering, Tennessee Tech University, Cookeville, TN, USA}
		\\\IEEEauthorblockA{\IEEEauthorrefmark{2}Electrical and Computer
Engineering, Texas A\&M University at Qatar, Doha, Qatar}\\	
		\IEEEauthorblockA{\IEEEauthorrefmark{3}Department of Electrical and Computer Engineering, Florida International University, FL, USA}\\	
	 \vspace{-8mm}}
	\maketitle
	\thispagestyle{empty}
	\begin{abstract}
Energy storage units (ESUs) enable several attractive features of modern smart grids such as enhanced grid resilience, effective demand response, and reduced bills. However, uncoordinated charging of ESUs stresses the power system and can lead to a blackout. On the other hand, existing charging coordination mechanisms suffer from several limitations. First, the need for a central charging coordinator (CC) presents a single point of failure that jeopardizes the effectiveness of the charging coordination. Second, a transparent charging coordination mechanism does not exist where users are not aware whether the CC is honest or not in coordination charging requests among them in a fair way. Third, existing mechanisms overlook the privacy concerns of the involved customers. To address these limitations, in this paper, we leverage the blockchain and smart contracts to build a decentralized charging coordination mechanism without the need for a centralized charging coordinator. First ESUs should use tokens for anonymously authenticate themselves to the blockchain. Then each ESU sends a charging request that contains its State-of-Charge (SoC), Time-to-complete-charge (TCC) and amount of required charging to the smart contract address on the blockchain. The smart contract will then run the charging coordination mechanism in a self-executed manner such that ESUs with the highest priorities are charged in the present time slot while charging requests of lower priority ESUs are deferred to future time slots. In this way, each ESU can make sure that charging schedules are computed correctly. Finally, we have implemented the proposed mechanism on the Ethereum test-bed blockchain, and our analysis shows that execution cost can be acceptable in terms of gas consumption while enabling decentralized charging coordination with increased transparency, reliability, and privacy preserving.      	

	\end{abstract}
	\begin{IEEEkeywords}
Blockchain;	decentralized system; charging coordination; energy storage units.

	\end{IEEEkeywords}

\section{Introduction}

Energy storage units (ESUs), including both home batteries and electric vehicles (EVs), present an effective mean to enhance the functionalities of the aging power grid. Specifically, ESUs represent a powerful emergency backup that can be used during electricity outage events, which in turn enhances the grid resilience~\cite{wang2018spatio}. Moreover, ESUs provide an approach to overcome the intermittent nature of renewable energy sources, which allows for high integration level of eco-friendly energy sources. Consequently, ESUs offer environmental benefits as well by reducing the greenhouse gas emissions. In addition, the stored energy in such units can be used during peak load periods, which in turn reduces the stress on the power grid during these periods, and hence, enables effective demand response. Furthermore, ESUs offer economic benefits by reducing the customers' electricity bills as the ESU owners can purchase energy from the grid during low tariff periods and then use it during high tariff periods. 

Despite their several benefits, ESUs pose several challenges that need to be addressed for efficient integration in the power grid. In specific, a simultaneous mass scale uncoordinated charging of ESUs may lead to lack of balance between the charging demands and the energy supply resulting in instability of the overall resilience of the grid~\cite{sortomme2011coordinated}. In severe cases, this could lead to a mass blackout. In order to mitigate such consequences, there is a substantial need for a charging coordination mechanism to avoid stressing the distribution system and prevent power outage~\cite{6204213}. In a charging coordination mechanism, ESUs should report data such as the time-to-complete-charging (TCC), the battery state-of-charge (SoC), and the amount of required charging. Then, then a trusted party i.e a charging controller define ESUs with highest priorities to charge and defer others to another time slot.

However, the existing charging coordination mechanisms\cite{wang2014semi, wang2018spatio} suffer from several limitations. First, the existing mechanisms rely on a single entity, namely charging coordinator (CC), to coordinate the charging requests. In turn, this can lead to the single server problem, i.e., if a successful denial of service attack is launched on the CC, it will be down and consequently a large number of charging requests could not be coordinated. Second, most of the existing works consider that the CC is a trusted party which is completely honest in scheduling charging requests. As a result, the ESU owners are not aware whether the charging schedules are computed correctly. In specific, a transparent charging coordination mechanism does not exist. Third, the existing charging coordination mechanisms require that the ESUs report some data to the CC such as whether an ESU needs to charge or not, the TCC, the battery SoC, and the amount of required charging. This,  in turn, reveals private information about the owners of the ESUs, such as the location of an EV and the activities of a house's residents \cite{a7,o4}. For example, the charging demands sent from an EV can reveal whether the EV's owner is at home and how long he/she will stay, and how often he/she drives. Also, if a home battery is not charged for an extended period, this can reveal that the residents do not spend time at home because they are traveling. The aforementioned limitations highlight the need for a decentralized, transparent, and privacy-preserving charging coordination mechanism.  

Recently, blockchains have attracted the attention of both academia and industry across a wide range of fields. With a blockchain in place, applications can operate in a decentralized and transparent fashion, without the need for a central authority, while achieving the same functionality  previously attained through a trusted intermediary. In addition, blockchains open the way to create smart contracts, which represent a piece of code on the blockchain that performs an action once certain criteria are met. Since it resides on the blockhain, smart contracts can be self-executed without the need for third parties~\cite{fernandez2018review}. Furthermore, the participants in a blockchain can be represented by anonymous identities rather than their real identities, and hence, the privacy of the users and the data is well preserved. Hence, blockchains can enable decentralized, transparent, and privacy-preserving charging coordination mechanism.\\ 
\textbf{Our contributions.} In this paper, we propose a blockchain-based charging coordination mechanism. A smart contract defining the rules to coordinate charging between different ESUs is proposed. Each ESU sends a charging request containing its power demand, SoC, and TCC to the smart contract address.  To preserve privacy, each ESU has a number of \textit{certified pseudonyms} and each pseudonym is used only for one charging request. The charging request is signed and this anonymous signature can prevent external attackers from sending valid charging requests. Based on the submitted data and the maximum available grid capacity, the smart contract will automatically determine a priority index for each ESU, and ESUs with the highest priority can be charged using a Knapsack algorithm~\cite{R}. To be the best of our knowledge, \textit{this work is the first to employ the blockchain technology to enable decentralized charging coordination mechanism} with the following features:
\begin{enumerate}
\item \textit{Reliability.} The scheme is reliable because of running the charging coordination mechanism does not rely on a single server (CC) and all the computations are carried out in a decentralized manner through the blockchain nodes~\cite{wood2014Ethereum}.
\item \textit{Privacy-preservation.} Since the ESUs use anonymous credentials, no one can link the SoC and TCC to an ESU that sent them. 
\item \textit{Transparency and verifiability.}  In our scheme, ESUs can run knapsack algorithm to ensure that the charging schedules are calculated correctly. 
\item \textit{Data integrity.} Once ESUs send their charging requests with TCC and SoC to the blockchain, they can later ensure the integrity of the requests' data since they have access to the blockchain, which cannot be done in centralized approaches.  
\end{enumerate}
The rest of this paper is organized as follows. In Section~\ref{prem} we discuss some preliminaries. In Section~\ref{network model}, we present the system model under consideration and adversary assumptions. Then, we introduce a temporal charging coordination approach in Section~\ref{knapsack_algorithm}. The blockchain-based privacy-preserving charging coordination mechanism is presented in section~\ref{proposedscheme}. The evaluations of the proposed mechanism are provided in Section~\ref{PerformanceEvaluation}. Finally, Section~\ref{conclusion} concludes the paper.

\section{Preliminaries}
\label{prem}

\subsection{Blockchain and smart Contracts}
\label{smart}

Blockchain  serves  as a fundamental structure of emerging cryptocurrencies such as Bitcoin~\cite{nakamoto2008bitcoin} to help make peer-to-peer exchange of value without a centralized third party. A blockchain is a distributed, immutable, and append-only data structure  formed  by  a  sequence  of  blocks  that  are  chronologically  and  cryptographically linked  together. The network is composed of a set of nodes called miners or validators are responsible of keeping a trustworthy-record  of  all  transactions  through  a consensus  algorithm in a trust-less environment~\cite{baza2018blockchain}. More importantly, blockchain enable the essence of smart contracts which can be defined as programs that every blockchain node will run them and update their local replicas according to the execution results without fraud or any interference from a third party. A blockchain presents two key elements, namely, transaction  and  addresses. Transaction is used  to modify or update the state stored in the blockchain network. Blockchain address represents the sender of a message in the blockchain is referred to a pseudonym. In practice, a blockchain address is usually bound to the hash of a public key\footnote{Through the paper, we use term public key and address interchangeably.}; more importantly, the security of digital signatures can further ensure that one cannot send messages in  the name  of  a  blockchain  address, unless  she has  the  corresponding  secret  key. Also,  the  program code of a smart contract deployed in the blockchain can also be referred  by  a  unique  address,  such that  one  can  call  the contract to be executed, by committing a message pointing to this unique address.

\subsection{Blinded and Partially Blinded signatures}
Blind signatures and Partial Blind Signatures (PBS) have been extensively used in the anonymization of electronic coins, and were introduced by~\cite{chaum1983blind}. Indeed, Blind Signature [14] allows the  sender of a message to obtain  signature  on  this message  from another party while concealing the content of the message. The signature requester generates a secret pair of blinding/unblinding operations $(b, b^{-1})$, and applies the blinding operation $b$ to a plaintext message $m$. Then, he/she sends the blinded message $b(m)$ to the signer who signs the blinded message with operation $s$ and produces a signature $s(b(m))$, and returns the signature to the requester. The requester uses the unblinding operation $b^{-1}$ to the signature to obtain $b^{-1}(s(b(m))) = s(m)$, that is a signature on the plaintext message. Partial Blind Signature (PBS) on the other side is a special case of Blind Signature where the signer can include a common message $m_0$ that is known to both singer and sender, such as a time or date. The requester submits blinded message $b(m)$, the signer generates signature $s(b(m),m_0)$, and returns it back to the requester. The requester applies the unblinding operation to get $b^{-1}(s(b(m),m_0)) = s(m,m_0)$. Note that the signer only knows the blinded message $b(m)$, not the plaintext message $m$. While the requester can unblind a signature $s(b(m),m_0)$ to obtain $s(m,m_0)$. The requester can further verify that $s(m,m_0)$ is indeed a valid signature on $(m,m_0)$, but cannot forge such a signature. Also, the signer can not link $s(m)$ to $b(m)$ and by this way the signer can not know the requester that requested the signature.

 \section{System and Threat models}
 In this section, we introduce the considered system model followed by adversary assumptions.
 
\label{network model}
\subsection{System model}
\begin{figure} [!t]

		\includegraphics[width=\linewidth]{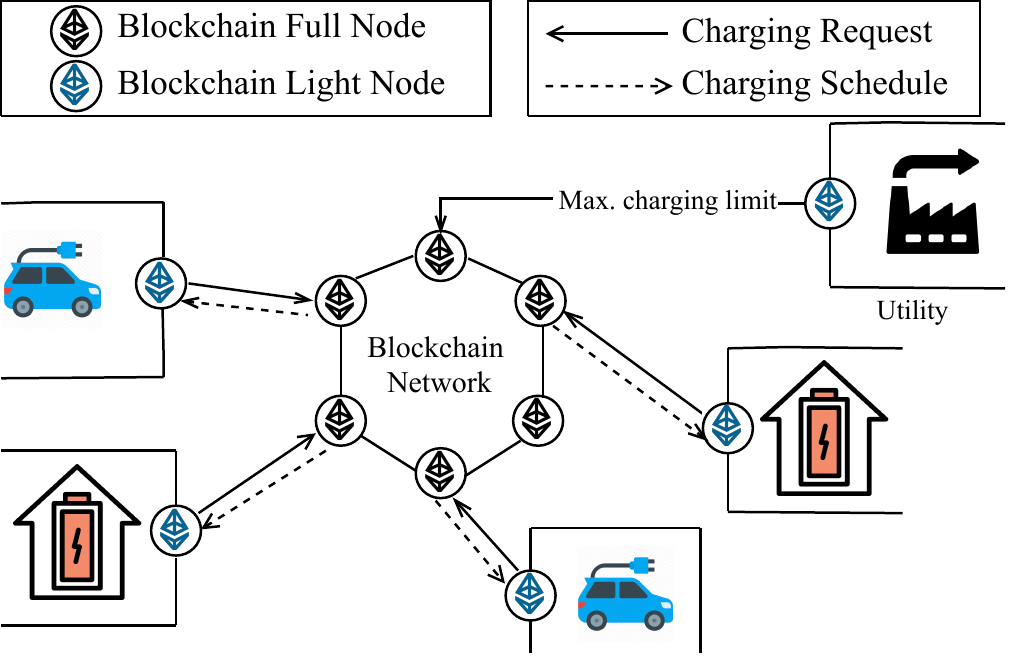}
	\caption{Illustration for the system model under consideration.} 
	\label{model}
\end{figure}

As depicted in Fig.~\ref{model}, our scheme has three main entities: Blockchain network, Energy Storage Units (ESUs.), and the utility.

\vspace{6pt}
\noindent \textit{Blockchain network:} is responsible to receive charging requests from ESUs and the total power load from the utility and schedule charging requests in a decentralized and transparent manner. The blockchain should support the smart contracts that is described in \ref{smart} so once requests are received, the code (smart contract) is executed  automatically and independently on each node of the blockchain network. 

\vspace{6pt}
\noindent \textit{Energy Storage Units (ESUs.)} ESU corresponds to the owner of the ESU (i.e,  electric vehicle and batteries) that can interact with the system through a mobile application. Each ESU should acquire a list of certified pseudonyms before sending charging requests to the blockchain. Also, as shown in Fig.~\ref{model}, it is not required that ESUs store a complete copy of the blockchain. Alternatively, they can be a blockchain light node by operating on Ethereum light client mode that allows lightweight devices such as Raspberry Pi devices to join the network, download block headers as they appear, and only validate certain pieces of state on-demand as required by their users.

\vspace{6pt}
\noindent \textit{The utility.} is responsible for posting the maximum load profile that ESUs in each community can charge and not coordinating the charging requests from ESUs.

\subsection{Adversary Model}

We follow the \textit{standard blockchain threat model} in~\cite{kosba2016hawk}, blockchain in our proposed design is maintained by a set of validators/miners, and is trusted for execution correctness of the proposed protocol and availability, but not for privacy. In this paper, we assume each ESU should have multiple \textit{certified pseudonyms}. In practice, a blockchain address is the hash of the public key and due to the security of digital signatures, only the owner of the public key can send messages in the name of its own blockchain address. In particular the attacker cannot reverse a one-way hash or forge digital signatures without the private key. We assume there are global eavesdropper including the utility that aim to passively collect some sensitive information about the ESUs owner's to guess the locations of the ESU owners. For example, learning about the current location of the EVs (if they are at home or not) with less than 30\% battery. If the ESU uses its true identity (e.g., long-term public key) to authenticate and send the charging requests, the utility will be able to know the ESU's location at a particular time, which compromises the ESU's privacy.

\section{Temporal Coordination of ESUs' Charging} \label{knapsack_algorithm}

This section describes a greedy algorithm that efficiently schedules the charging of the ESUs. In the next section, we will discuss how to implement this scheduling algorithm as a smart contract in the blockchain.

The ESUs are deployed within a set of communities. Each community is associated with an electric bus that presents a loading limit of $C$ kW. The regular load profile capacity within a specific community can be denoted by $P_R$ kW at a given time slot, which accounts for residential, commercial, and industrial loads. 
In some cases, it is not possible that all ESUs with charging requests can be scheduled for charging at the present time slot due to the capacity limitation. Hence,  a priority index is calculated for each ESU and the ESUs with the highest priorities will be charged at the present time slot while ensuring that $\sum P_v \leq C - P_R$, where $P_v$ is the amount of charging demand requested by an ESU . Meanwhile, other ESUs' charging requests can be deferred to future time slots. 

To determine the ESU's priority for charging at the present time slot, two main components play a vital role, namely, the TCC ($K_v$) and the battery SoC ($S_v$). Typically, an ESU with low $S_v$ and short $K_v$ should have higher charging priority than an ESU with high $S_v$ and/or potentially long $K_v$. Subsequently, for each ESU $v$, we specify a priority function $U_v$ such that 
\refstepcounter{defcounter}
\renewcommand\theequation{\thedefcounter}
\begin{equation}
    U_v = \beta_1 (1 - S_v) + \beta_2 F(K_v),
     \label{1}
\end{equation}
where $\beta_1$ and $\beta_2$ can be used as  weights to determine the relative significance of $S_v$ and $K_v$, with $\beta_1$ $+$ $\beta_2$ $=1$, $F(K_v)$ is a decreasing function of $K_v$ with a range of $[0,1]$ and $F(K_v) = 0$ for long TCC and equals $1$ for short TCC, and SoC value ($S_v) \in [0,1)$ with $S_v = 1$ for a completely charged ESU. To illustrate, an ESU $v$ with low $S_v$ value and short $K_v$ will have a high priority value $U_v$. Then, because of the limiting capacity in the present time slot, our goal is to schedule the ESUs that have the highest priority for charging in the present time slot, and postpone the charging of ESUs that have lower priorities to future time slots.

The charging schedule indicates whether a given ESU $v$ will be charged in the present time slot ($x_v = 1$) with the charging request ($P_v$) or not ($x_v = 0$), where
\refstepcounter{defcounter}
\renewcommand\theequation{\thedefcounter}
\begin{eqnarray}
\begin{aligned}
 & \underset{x_{v} \in \{0, 1\}}{\max}
 & &  \sum_{v \in \mathcal{V}} x_v U_{v}\\
 & s.t.
 & & \sum_{v \in \mathcal{V}} x_v P_v \leq C - P_R.
     \label{3}
\end{aligned}
\end{eqnarray}
It should be noted that a scheduled ESU can obtain its entire charging demand ($P_v$) at the current time slot according to (\ref{3}). To ensure an efficient resource utilization, the remaining capacity after the scheduling of ESUs according to (\ref{3}), $C-P_R - \sum x_v P_v$, will be assigned to the ESU with highest priority among all unscheduled ESUs. Such an ESU which has not fully charged at the current time slot, will be suspended to the next slot to obtain whatever is left of its charging request.

The charging coordination problem in \eqref{3} at each time slot can be well described as a Knapsack problem \cite{R} where a set of items with different values and weights need to be packed in a knapsack of maximum capacity. The Knapsack problem aims to select the items with the highest values while satisfying the knapsack capacity constraint. In this context, the ESUs are mapped to the items, the ESU priority $U_v$ represents the item value, the ESU charging request $P_v$ resembles to the item weight, and the charging limit constraint $C-P_R$ is equivalent to the knapsack limit. Hence, in order to to schedule ESU charging at each time slot according to \eqref{3}, we have modified a greedy algorithm for solving the Knapsack problem in polynomial time complexity \cite{R}. The greedy algorithm that solves \eqref{3} sorts the ESU in a descending order according to the ratio between their priorities and weights. Then, ESUs with the highest orders are scheduled for charging in the current time slot, while accounting for the capacity limit.  

\IncMargin{1em}
\definecolor{seagreen}{rgb}{0.18, 0.55, 0.34}
\SetAlFnt{\small\color{black}\tt}
\LinesNumbered
\renewcommand{\DataSty}[1]{{\color{yellow}\texttt{#1}}}
\renewcommand{\KwSty}[1]{{\color{blue}\texttt{#1}}}
\renewcommand{\ProgSty}[1]{{\color{black}\texttt{#1}}}
\renewcommand{\FuncSty}[1]{{\color{black}\texttt{#1}}}
\renewcommand{\CommentSty}[1]{{\color{seagreen}\texttt{#1}}}
\renewcommand{\ArgSty}[1]{{\color{black}\texttt{#1}}}
\renewcommand{\FuncArgSty}[1]{{\color{black}\texttt{#1}}}
\renewcommand{\ProcArgSty}[1]{{\color{black}\texttt{#1}}}
\renewcommand{\ProcNameSty}[1]{{\color{black}\texttt{#1}}}
\renewcommand{\NlSty}[1]{{\color{black}\texttt{#1}}}

\begin{algorithm*}[!t]
	\SetKwProg{Fn}{function}{}{}
	\SetKwProg{St}{Struct}{}{}
	\SetKwProg{ma}{mapping}{}{}
	\SetKwProg{Contract}{contract}{}{}
	\SetKwData{NumOfUpdatedObjects}{numOfUpdatedObjects}
	\SetKwIF{If}{ElseIf}{Else}{if}{}{else if}{else}{end if}
	\SetKwFunction{ChargingCoordination}{Charging\_Coordination}
	\SetKwFunction{CalculatePriority}{Calculate\_Priority}
	\SetKwFunction{RecievechargeRequest}{Recieve\_Charging\_Request}
	\SetKwFunction{Knapsack}{Knapsack}
	\SetKwFunction{QuickSort}{Quick\_Sort}
	\Contract{\ChargingCoordination}{
	  \BlankLine
	  \Fn{\ChargingCoordination{\_owner, PR}}{ 
		  
		  owner $\leftarrow$ \_owner\\
		  
		  MaxCapacity $\leftarrow$ C - PR \tcp{Charging Capacity}
	  }
	  \BlankLine 
	   \St{ESU{(Pv, TTC, SoC, Priority, xv, PScuduled)}}{}
	   \ma{(\textcolor{blue}{address} => ESU) ESUs}{}
	   \textcolor{blue}{address} [] ESUlist 

	\Fn{\RecievechargeRequest{\_Pv, \_TTC, \_SoC, $\sigma(ID_v)$}}{  \tcp{receive charging requests}
		  
		\If{Is\_Authorized(msg.sender,$\sigma(ID_v)$)}{
			ESUlist.push(msg.sender)\\
			ESU.Pv $\leftarrow$ \_Pv   \\
			ESU.TTC $\leftarrow$ \_TTC \\
			ESU.SoC $\leftarrow$ \_SoC \\
			ESU.Priority $\leftarrow$   ( ( ( 500 * 1- \_TTC ) + ( 500 * \_SoC ) ) / \_Pv );
		}
		
	}

	  \BlankLine
	\Fn{\Knapsack{}}
	{
		  QuickSort(0,ESUlist.length-1);\tcp{call QuickSort} 
		  \For{i $\leftarrow $ 0  to ESUlist.length-1}
		  { 
		  \If{ESUs[ESU\_list[(i)]].Pv <= MaxCapacity}
			{
				ESUs[ESU\_list[i]].xv $\leftarrow$ 1 \\
		  ESUs[ESU\_list[i]].PScuduled $\leftarrow$ ESUs[ESU\_list[i]].Pv\\
			 MaxCapacity  $\leftarrow$ MaxCapacity-ESUs[ESU\_list[i]].Pv\\
			}
		}
	 }      
	  \BlankLine

		\Fn{\QuickSort{\_Start, \_End}}{ 
			Start $\leftarrow$ \_Start\\
			End $\leftarrow$\_End\\
			\lIf {Start = End}{\Return}
			pivot $\leftarrow$ ESUs[ESU\_list[(Start + (End - Start) / 2)]].Priority;\\
			\While{Start <= End }{ 
				\While{ESUs[ESU\_list[Start]].Priority > pivot) }{Start++}
				\While{pivot > ESUs[ESU\_list[End]].Priority}{End\texttt{-{}-}}
				\If{Start <= End}{ 
					   (ESU\_list[Start], ESUlist[End]) $\leftarrow$ (ESU\_list[(End)], ESUlist[Start])\\
					   Start++  \\ 
						End\texttt{-{}-} \\
				}
			}
			\lIf{\_Start < End}{QuickSort(\_Start, End)}
			\lIf{Start <  \_End}{QuickSort(Start, \_End)}

		} 
	}
\caption{Pseudocode for the charging coordination contract}
\label{alg:contract}
\end{algorithm*}

\section{Blockchain-based Charging Coordination}
\label{proposedscheme}
This section describes the proposed blockchain-based charging coordination mechanism. The proposed mechanism operates in four phases, namely, acquiring anonymous credentials, charging request submission, charging coordination. The charging coordination algorithm described in Section \ref{knapsack_algorithm} is implemented via a smart contract described in the pseudo code of Algorithm~\ref{alg:contract}. Two data types are supported in our smart contract, namely, address and mapping. The address is a special data type that is used to store the message caller, and the mapping resembles a hashmap that stores each ESU-related data such as TCC, SoC, and the charging amount. The smart contract described in Algorithm~\ref{alg:contract} consists of a constructor named \texttt{\texttt{Charging\_Coordination}} and the following methods (functions), namely \texttt{Recieve\_Charging\_Request}, \texttt{Knapsack}, and \texttt{Quick\_Sort}.

\vspace{-2mm}
\subsection{Acquiring Anonymous Credentials}
In order to allow the smart contract to authenticate charging requests for a group of ESUs anonymously, each ESU should request a Partial Blind Signature (PBS)  (e.g., \cite{okamoto2006efficient}) from the utility on public key that will be used to generate its Etherum address during charging request submission phase as follows. 
Assume that the utility has a key pair $(P_\tau , S_\tau)$. Each ESU $(v)$ should acquire $N$ tokens $\tau_{i}: i=1,\cdots,N$. For each $i$, the ESU generates a random secret $x_i$ and computes its public key $PK^{i}_v$, Then, it blinds each $PK^{i}_v$ using $b_v$ and signs the message with its true identity (e.g., using ECDSA), and sends: $b_{v}(PK^{i}_v),\sigma_v$ to the utility, where $b_{v}(PK^{i}_v)$ is the blinded public key and $\sigma_v$ is ESU $v$ digital signature on the entire message. The utility generates a partially blind signature $PBS^{m_0}_{S_\tau}(PK^{i}_v)$ where $m_0=TS||ID_g$ is the appended common message, which is the current date and identifier of the group or community $ID_g$ that the ESU belongs to. The utility then returns $PBS^{m_0}_{S_\tau}(PK^{i}_v), \sigma_U$
to the ESU, where $\sigma_U$ is the utility's digital signature on the entire message. Then, the ESU verifies $\sigma_U$, and applies the unblinding operation $b^{-1}_v$ to obtain the token 
$$\tau_{i}=b^{-1}_v(PBS^{m_0}_{S_\tau}(b_v(PK^{i}_v)) = PBS^{m_0}_{S_\tau}(PK^{i}_v),$$
and verifies $\tau_{i}$ is a valid signature on $(PK^{i}_v)$ and $m_0$ using the public key $P_\tau$. Note that in this phase, although the ESU uses its real identity, the utility can not know the contents of the message and this phase can occur every long period like a week. 
\vspace{-2mm}
\subsection{Charging Request Submission}

In this phase, the ESUs submit a charging request ($R_v=P_{v}, S_{v}, K_{v}, TS, ID_g, \sigma(PK_v|TS|ID_g)$) to the blockchain. The charging request is made as a transaction submitted to the smart contract address on the blockchain, and cannot be linked directly to a specific ESU. As a fundamental concept in blockchain, the ESU address changes for each transaction for privacy preservation reasons. The method \texttt{Recieve\_Charging\_Request} can be called by an ESU to submit a charging request to the smart contract. This method calls  \texttt{Is\_Authorized} method that checks if the sender address and the anonymous signature $\sigma$ are from a legitimate ESU, in such case the sender address and the request info are added to the requests list of the current time slot. Since, Ethereum does not support fixed point numbers and in order to allow our contract to compute the priority index described in Equation~\ref{1}, we have mathematically represented $\beta_{1}$ and $\beta_{2}$ as numbers in the range of [0, 1000]. Then, the ESU's priority value is divided by the requested charging amount as shown in Algorithm \ref{alg:contract}.

\vspace{-2mm}
\DecMargin{1em}
\subsection{Charging Coordination}

    



By the end of each time slot, a greedy algorithm for solving the Knapsack problem is executed over the received charging requests by triggering the method \texttt{Knapsack}, which first calls the \texttt{QuickSort} method that sorts the ESUs in a descending order according to the ratio of the priority index to the amount of requested charging. Finally, each ESU will be assigned a certain charging amount at a given time slot.

It is worth mentioning that smart contracts need to be triggered by an external account to do some operation. In order to ensure that the \texttt{Knapsack} method is executed at the end of each time slot, Aion\footnote{https://github.com/ETH-Pantheon/Aion}, which is a smart-contract-based system deployed at address 0xCBe7AB529A147149b1CF982C3a169f728bC0C3CA at the Ethereum blockchain, is employed. With Aion, transactions of any type such as contract's function executions can be scheduled to be executed at specific time instants. In other words, the charging coordination process can be carried out in a completely automatic way (at each time slot) without any interference from any other party.

\section{Performance Evaluation}
\label{PerformanceEvaluation}
\subsection{Charging Coordination Evaluation}

\begin{figure} [!t]
		\includegraphics[width=\linewidth]{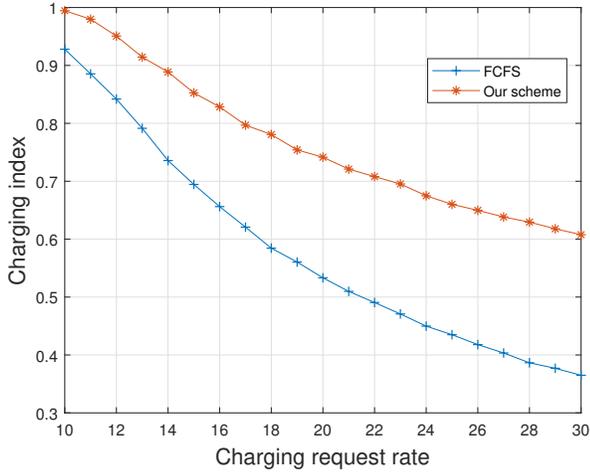}
	\caption{Average charging index versus charging request rate.}
	\label{C}
\end{figure}

First, we evaluate the proposed charging coordination mechanism by comparing it with the First-Come-First-Serve (FCFS) approach, in which the ESU that demands charging first gets charged regardless of its TCC or SoC. We conduct the experiment for $30$ time slots with a battery capacity of $200$ kW and maximum capacity for charging per time slot ($C-P_R$) of $1000$ kW.  First, there are 10 ESUs that need to charge, and a Poisson distribution with an average of $\lambda$ is used to simulate the arrival process of new charging requests at each time slot. The battery SoC is a random number in the range of $[0, 1]$  that follows  a uniform distribution, while the TCC is a random number that follows a Geometric distribution with an average of $4$. The priority function, $F(K_v)$ is set to 1, 0.5, and 0, when $K_v$=1, $K_v$=2, and $K_v$ > 3 respectively. The results  are the average of $80$ runs. 
We use the charging index as a metric for performance evaluation. It is calculated by dividing the amount of power an ESU charges by the amount of required charging, which is a value in the range $[0,1]$. 
Fig.~\ref{C} shows the charging index of our mechanism at different values of $\lambda$ compared with FCFS. The figure indicates that our mechanism can deliver higher charging power to the ESUs.

\subsection{Computation Cost Analysis}
\begin{table}[!t]
\centering
\begin{tabular}{lll}\toprule
        &   GAS  &  Price (ETH) \\ \midrule
\texttt{Deploy\_Contract}  &742276             &0.0044537 \\

\texttt{Recieve\_Charging\_Request} &76981           & 0.0004619   \\
\bottomrule
\end{tabular}
\caption{Summary of execution costs.}
\label{cost1}
\end{table} 

The Ethereum blockchain uses the \textit{Ethereum Virtual Machine} (EVM) to execute the code of smart contracts. This EVM is quasi-Turing complete. A machine is turing complete when it is able to solve any calculable problem given enough space and time, and it is quasi as the EVM requires enough GAS (Ether) units to operate. Every instruction consumes GAS units to be executed in the EVM. For example: to add two values from memory, 3 GAS units should be paid ~\cite{wood2014Ethereum}. If a user wants to execute a smart contract and sends a transaction to the Ethereum network that contains the instruction to do something, the user has to pay GAS units.

We have implemented a smart contract for Algorithm \ref{alg:contract} in Solidity 0.4.0, which allows to design such contracts with private and public methods and has a set of basic data types. The smart contract was deployed into the Kovan blockchain ~\cite{kovan} in block 24,538 with address 0xf87c410e1b35a4424e764873964220818a222993. 

\begin{figure} [!t]
		\includegraphics[width=\linewidth]{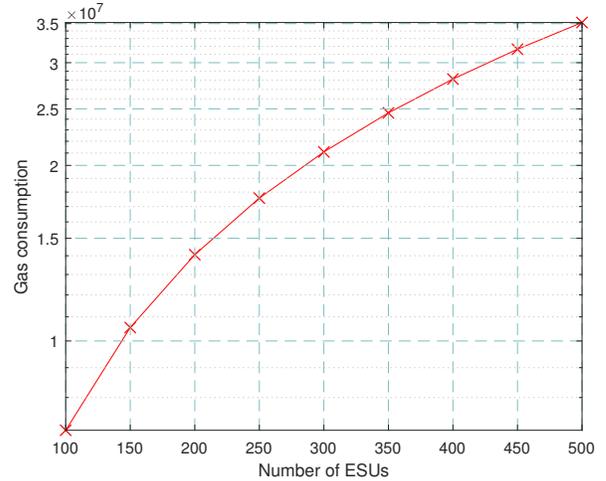}
	\caption{Execution cost of \texttt{Knapsack} method.}
	\label{gas}
\end{figure}

The execution costs of our mechanism can be considered as follows. The execution cost of deploying the smart contract to the blockchain, the cost of calling \texttt{Recieve\_Charging\_Request} method by an ESU, and the cost of running the \texttt{Knapsack} method.  According to \cite{gasstation}, the cost for 1 unit of GAS is on average 5 Gwei = $5 \times 10^9$ ETH. Table~\ref{cost1} shows the execution cost of deploying the contract and \texttt{Recieve\_Charging\_Request} method. It can be noted that the costs are relatively low. For the execution cost of the \texttt{Knapsack} method, Fig.~\ref{gas} shows that as the number of ESUs sending charging requests increases, the GAS consumption increases. Despite the reduced complexity of the proposed coordination algorithm, using the \texttt{Quick\_Sort} method presents a complexity of $|\mathcal{V}| \log |\mathcal{V}|$ that incurs an increase in the expected cost. While smart contracts offer a very promising way to implement protocols in a privacy-preserving as well as transparent manner without the need to rely on a single centralized coordinator, it is still in its early evolving stage, and some limitations need to be addressed such as the scalability of number of operations made by the EVM.      

\vspace{-2mm}	
\subsection{Security/Privacy Analysis}
Besides the fair charging coordination feature that stemmed from the Knapsack algorithm, our proposed mechanism presents the following unique features:

\begin{enumerate}
\item \textit{Decentralized charging coordination.} Since the blockchain is responsible for executing the charging coordination among interested parties, it is impossible to have a single point of failure. An attacker needs to control a massive number of blockchain nodes in order to reach a system failure state, which is practically impossible.

\item \textit{Privacy-preserving charging activities.} In the proposed mechanism, the privacy of ESU owners (their charging requests including the SoC, TCC) is protected by $(i)$ replacing the ESU's real identities by some placeholders (pseudonyms) for charging requests that corresponds to temporary public-private key pairs, and $(ii)$ the anonymous credentials which are obtained using PBS without allowing the utility to link them with the the true ESUs identities. Every pseudonym expires once the ESU owner send a charging request to the blockchain which ensures \textit{unlinkabilty}. Also, because of anonymous authentication done by the blockchain, external attackers can not damage the scheme by sending charging requests data while they do not belong to a specific community.

\item \textit{Availability.} the proposed scheme resists against Denial-of-service (DoS) attacks. In such an attack, an attacker targets an ESU or even the utility to prevent legitimate transactions from appearing on the ledger, thus preventing them from posting new charging requests. To launch this attack successfully, the attacker will need to control the majority of the mining power of the network, which is practically impossible.


\item \textit{Data integrity and transparency.} Since each user has an access to the blockchain, he/she can verify the charging request sent by him and more importantly, the results of running the modified Knapsack algorithm can be verified so that the user can check whether he/she has a priority to charge or not. As a result, the proposed mechanism offers high transparency which is not offered in case of using centralized approaches.
\end{enumerate}

\section{Related Work}

In ~\cite{mahmoud2016privacy}, a privacy a ware charging coordination mechanism is proposed. Each ESU should send a charging request to an aggregator. The aggregator forwards the requests to a charging controller to run a charging coordination
mechanism to define ESUs with height priorities. altthough With using several  neither the aggregator nor the CC can have access to the ESUs' sensitive information However, the proposed scheme does not provide transparency to users. 

Recently, blockchain has been considered as one of the rising technologies that can be used to adopt blockchain based applications for smart grids. Inspired by bitcoin, a PriWatt system is introduced by ~\cite{aitzhan2018security} that enables a blockchain-based private decentralized energy trading system. The system allows peer-to-peer energy trading without the need for a third-party intermediary.

In~\cite{gao2018gridmonitoring}, a smart grid monitoring system is proposed based on blockchain and smart contracts to ensure provenance, and immutability of smart metering data. In~\cite{knirsch2018privacy}, a blockchain-based dynamic pricing mechanism is proposed for EV charging. The mechanism allows customers to find the cheapest charging station within a previously defined region while preserving the privacy of the customers. Different stations store their bids for tariffs on the blockchain based on the requested energy. 
Unfortunately, the existing research works do not present a mechanism that enables decentralized, transparent, and privacy-preserving charging coordination, which schedules incoming charging requests in a manner that efficiently utilizes the power grid capacity while not stressing the grid.

		\section{CONCLUSION}
	\label{conclusion}
	In this paper, a charging coordination mechanism for ESUs has been proposed based on the blockchain technology. First, a temporal charging coordination mechanism is presented based on the Knapsack problem to maximize the power delivered to the ESUs while respecting the grid capacity limitations. Different from the traditional centralized implementation, a prototype implementation of the proposed coordination mechanism is deployed on the Ethereum blockchain so that ESUs can get their charging demands in a decentralized, transparent, and verifiable manner. 
	Finally, we found that the costs for deploying and executing the implemented smart contract in Ethereum is reasonable for a small scale community of ESUs. In our future work, we plan to implement our mechanism in a blockchain technology that does not rely on cryptocurrency to execute the application so that the coordination mechanism can be implemented in a large scale. 
	\bibliographystyle{IEEEtran}
	\bibliography{references}

\end{document}